\documentclass[twocolumn,showpacs,preprintnumbers,amsmath,amssymb,prb]{revtex4}

\usepackage{graphicx}
\usepackage{dcolumn}
\usepackage{bm}

\def\U#1{{%
\def\O{\mbox{O}}
\def\u{\mbox{u}}
\mathcode`\u=\mu
\mathcode`\O=\Omega
\mathrm{#1}}}

\def\ii{{\mathrm{i}}}
\def\dd{{\mathrm{d}}}
\def\sub#1{_{\scriptsize\mbox{#1}}}
\def\ee{{\mathrm{e}}}

\begin{document}

\preprint{}

\title{Electromagnetic response of a metamaterial with
field-gradient-induced transparency}

\author{Yasuhiro Tamayama}
\email{tama@giga.kuee.kyoto-u.ac.jp}
\author{Toshihiro Nakanishi}
\author{Yasuhiro Wakasa}
\author{Tetsuo Kanazawa}
\author{Kazuhiko Sugiyama}
\author{Masao Kitano}
 \affiliation{Department of Electronic Science and
 Engineering, Kyoto University, Kyoto 615-8510, Japan}

\date{\today}

\begin{abstract}

We investigate a dynamically controllable 
electromagnetically induced transparency-like metamaterial. 
The unit structure of the metamaterial consists 
of a low-quality-factor resonator and a high-quality-factor resonator. 
The field gradient of the incident electromagnetic wave in the 
transverse direction induces coupling between these two types of 
resonators and causes a transparency phenomenon. 
We present the simulation and experimental results for the 
dynamic control of the electromagnetic response.

\end{abstract}

\maketitle

\section{Introduction}

A number of studies have focused on controlling the propagation of
electromagnetic waves using metamaterials that consist of
periodically or randomly arranged artificial subwavelength structures. 
Metamaterials can even be used to obtain extraordinary media that do
not occur in nature, such as media with a negative
refractive index.\cite{shelby01}
Metamaterials have been applied to demonstrate 
various phenomena, such as 
subwavelength imaging,\cite{lagarkov04,liu07}
cloaking,\cite{schurig06,liu_r09} and 
no-reflection propagation.\cite{tamayama06,edwards08}

On the other hand, 
electromagnetically induced transparency (EIT) has attracted considerable
attention in recent years as a means by which to achieve extremely low group
velocity propagation of electromagnetic waves.\cite{harris97,fleischhauer05} 
Electromagnetically induced transparency is a quantum phenomenon that
arises in three-state $\Lambda$-type atoms 
interacting with electromagnetic fields, and 
causes the suppression of the absorption of incident electromagnetic waves in a
narrow frequency range. 
Based on the Kramers-Kronig relations,\cite{saleh07}
steep dispersion, or
low group velocity, can be achieved in the frequency range. 
In fact, a group velocity of $17\,\U{m/s} \sim 10^{-7} c_0$ 
(where $c_0$ is the speed of light in a vacuum)
has been observed using the phenomenon.\cite{hau99} 

Since a somewhat complicated arrangement is necessary in order to
achieve EIT, a number of studies have focused on mimicking 
the effect in classical systems, such as
optical resonators,\cite{smith04,yanik04_2,totsuka07,yang09} 
waveguides,\cite{liu_opex09} and 
metamaterials.\cite{zhang_prl08,papasimakis08,papasimakis09,tassin_prl09,tassin_opex09,liu_nat09,yannopapas09,chiam09} 
However, few studies have focused on the dynamic control of 
the transmission spectrum in classical EIT-like systems. 
In the present study, we investigate an EIT-like metamaterial having
properties that can be controlled dynamically. 

First, we describe the structure of the EIT-like metamaterial. 
The unit cell of the metamaterial consists of a low-quality-factor ($Q$)
resonator and a high-$Q$ resonator. 
The field gradient of the incident electromagnetic wave in the transverse direction
induces coupling between these two types of 
resonators and causes a transparency
phenomenon in the metamaterial. 
Then, we describe how to control the electromagnetic response of the
metamaterial. We present the experimental results for the dynamic
control of the transparent frequency using variable capacitance
diodes and the simulation results for the
dynamic control of the frequency width of the transparency window. 
Although the experiment is performed in the microwave region, 
it may be possible to perform similar experiments in the
terahertz and optical regions.

\section{Theory}

\begin{figure}[tb]
\begin{center}
\includegraphics[width=7cm,clip]{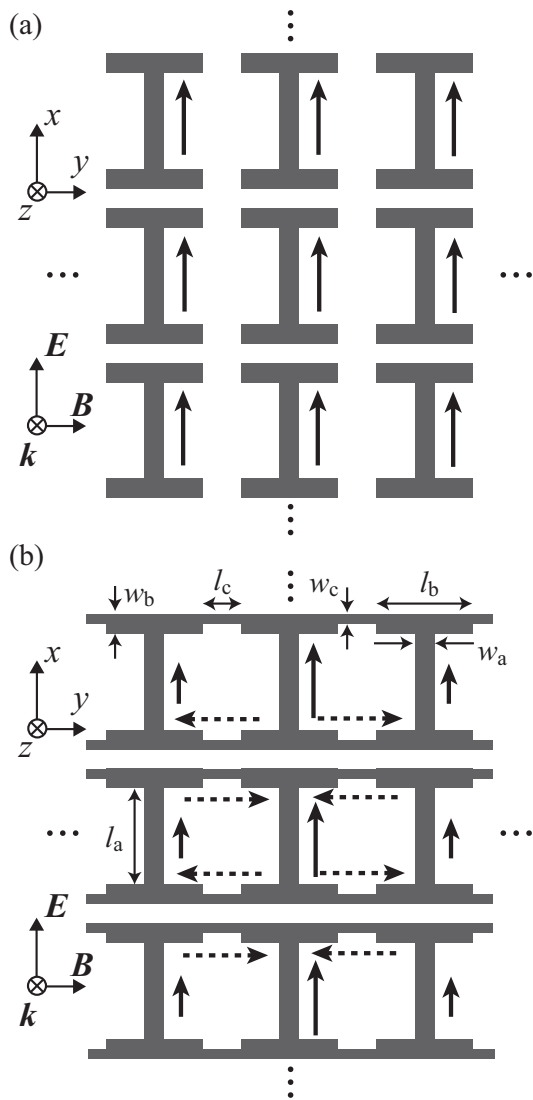}
\caption{(a) Metal structure of the capacitively loaded strips and 
(b) the field-gradient-induced-transparency metamaterial. 
The solid (dashed) arrows represent the current in resonator 1
 (resonator 2).}
\label{fig:1}
\end{center}
\end{figure}

Figure \ref{fig:1}(a) shows a capacitively loaded
strip metamaterial.\cite{ziolkowski03} 
The metamaterial can be regarded as series inductor-capacitor resonant 
circuits that consist of metal strips in the $x$ direction and
gaps between the neighboring structures in the $x$ direction. 
When a plane electromagnetic wave is incident on the metamaterial, 
as shown in Fig.\,\ref{fig:1}(a), 
the metamaterial behaves as a Lorentz-type dielectric medium because
currents are induced in the $x$ direction 
by the incident electric field and the electric dipoles are excited. 

We now consider the metamaterial shown in Fig.\,\ref{fig:1}(b). 
Each capacitively loaded strip is connected to the neighboring structures
in the $y$ direction. 
The metamaterial shown in Fig.\,\ref{fig:1}(b) includes two types of
series 
inductor-capacitor resonant circuits. One consists of a metal strip in the
$x$ direction and a gap between the neighboring structures in the
$x$ direction, which 
is the same as the resonant circuit in the capacitively loaded strips. 
The other consists of a metal strip in the $y$ direction and a 
gap between the neighboring structures in the $x$ direction. 
(The numerically calculated 
current and electric field distributions in the metamaterial are shown 
in Fig.\,\ref{fig:3-2} and can be used to determine the portions of each
structure that act as inductors and capacitors.) 
The former resonant circuit is referred to as resonator 1, and the
latter resonant circuit as resonator 2. These two types of resonators
are magnetically coupled 
to each other. The $x$-polarized incident 
electric field can directly excite only resonator 1. 
Resonator 2 cannot be excited directly because the direction of current
flow in resonator 2 is perpendicular to the direction of the incident electric field. 

Magnetic coupling occurs between the two types of resonators 
when the field gradient of the incident electromagnetic fields exists. 
When the $x$ component of the electric field has a 
gradient in the $y$ direction, the induced current in resonator 1 
becomes $y$ dependent. 
The difference between adjacent currents creates a magnetic flux through
the loop of resonator 2 
and induces anti-parallel currents via the electromotive force.
When the incident electromagnetic wave is a plane wave, this coupling
vanishes due to the uniformity in the $y$ direction of the currents in
resonator 1. Thus, only resonator 1 can be excited. In this case, the
metamaterial behaves as a Lorentz-type medium, as in the case of the
capacitively loaded strips. 

The resonance mode in resonator 1 is electric dipole resonance
whereas that in resonator 2 is weak radiative resonance, i.e., 
trapped-mode resonance.\cite{fedotov07} 
This implies that resonator 2 has a higher $Q$ value than resonator 1. 
Therefore, this system is similar to the classical model of EIT.\cite{alzar02} 
When the difference between the frequency of the incident 
electromagnetic wave and the resonance
frequency of resonator 2 is large, 
resonator 2 is excited only slightly and the properties of the metamaterial are
determined primarily by resonator 1. Thus, the metamaterial behaves almost as
a Lorentz-type dielectric medium. 
When the frequency of the electromagnetic wave is nearly equal to the
resonance frequency of resonator 2, 
resonator 2 is excited through the coupling with resonator 1. 
Resonator 2 stores the electromagnetic energy with low loss because its
$Q$ value is high.  
The current in resonator 1 is the sum of the current induced by the
incident electromagnetic wave and the current induced by resonator 2. 
Due to destructive interference
between these currents, the current in resonator 1 might vanish, at which point the metamaterial would become transparent. 
Since the transparency phenomenon is induced by the field gradient 
of electromagnetic waves, we refer to the metamaterial as a \textquotedblleft
field-gradient-induced-transparency 
metamaterial.\textquotedblright 

We can tune three parameters
in order to control the electromagnetic response of our 
metamaterial: the resonance frequencies of resonators 1 and 2, and the
field gradient in the $y$ direction of the incident electromagnetic wave.
The first two parameters determine the center frequency of the absorption
line and that of the transparency window, respectively. 
The third parameter corresponds to the coupling strength 
between resonators 1 and 2 and  
determines the frequency width of the transparency window.

\section{Experiment on dynamic control of transparent frequency}

\begin{figure}[tb]
\begin{center}
\includegraphics[width=7cm,clip]{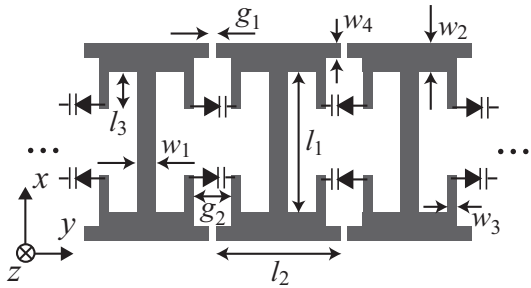}
\caption{Schematic of the metamaterial used for the experimental
 demonstration. 
The geometrical parameters are $l_1 = 29.1\,\U{mm}$, $l_2 =
 11.8\,\U{mm}$, $l_3 = 5.0\,\U{mm}$, $w_1 = 3.0\,\U{mm}$, $w_2 =
 2.0\,\U{mm}$, $w_3 = 0.5\,\U{mm}$, $w_4 = 1.0\,\U{mm}$, $g_1 =
 0.4\,\U{mm}$, and $g_2 = 1.0\,\U{mm}$. 
}
\label{fig:2}
\end{center}
\end{figure}

We fabricated the metamaterial shown in Fig.\,\ref{fig:2} 
using printed circuit board technology. 
The thicknesses of the copper layer and the FR-4 substrate (relative
permittivity: 4.5) of the printed circuit board were 
$35\,\U{um}$ and $1.6\,\U{mm}$, respectively. 
We used a structure that was slightly different from that shown in 
Fig.\,\ref{fig:1}(b) to 
adjust the resonance frequencies of resonators 1 and 2. 
We introduced variable capacitance diodes (Infineon BB857) for 
dynamic control of the resonance frequency of resonator 2. 
Their capacitances were tuned 
by applying a direct current reverse bias voltage.\cite{gil04} 

We measured the transmissivity and the group delay of the fabricated
metamaterial using a network analyzer. 
A layer of the metamaterial was placed in a rectangular waveguide, 
the cross-sectional dimensions in the $x$ and $y$ directions of which 
were $34.0\,\U{mm}$ (height) and $72.1\,\U{mm}$ (width), respectively. 
The measurement was performed in the frequency region 
in which only the TE$_{10}$ mode\cite{collin90} can propagate. 
The waveguide walls parallel to the $yz$ plane are equivalent to
periodic boundaries because the electromagnetic fields are uniform in
the $x$ direction. 
The width of the waveguide determines the field gradient of the
electromagnetic waves in the $y$ direction. 

\begin{figure}[tb]
\begin{center}
\includegraphics[width=7cm,clip]{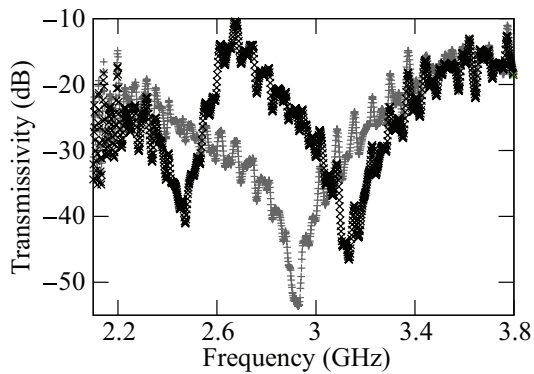}
\caption{Transmission spectra of the metamaterial shown in
 Fig.\,\ref{fig:2} with diodes (black) and without diodes (gray). 
The reverse bias voltage on the diodes is zero. 
}
\label{fig:3-1}
\end{center}
\end{figure}

\begin{figure*}[tb]
\begin{center}
\includegraphics[width=15cm,clip]{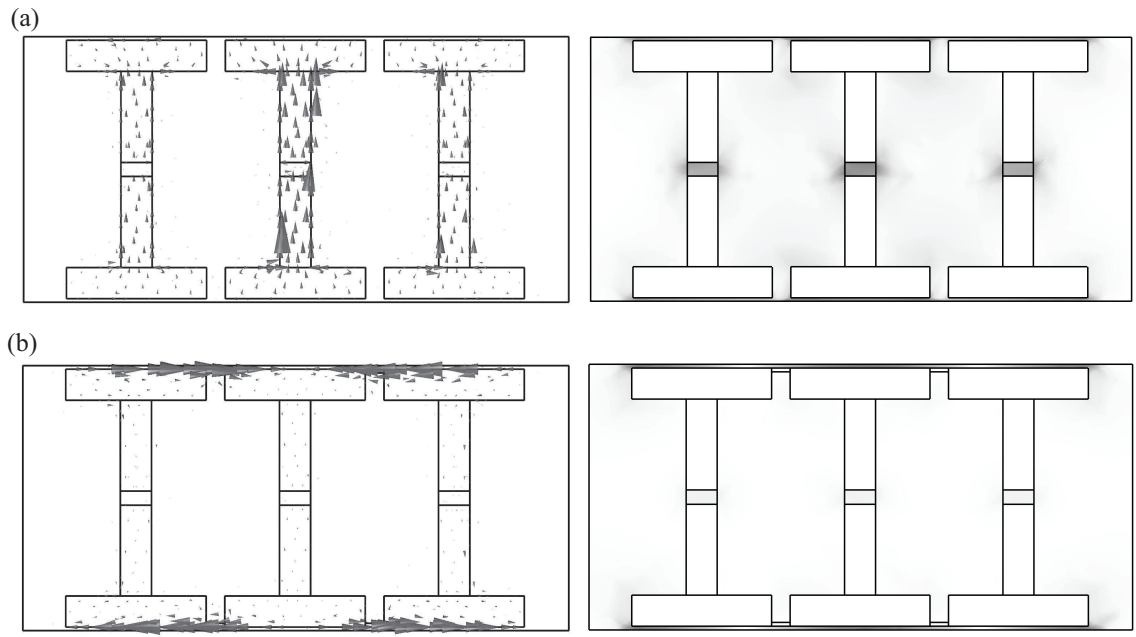}
\caption{Current distributions (left panel) and electric field
 distributions (right panel) in (a) the capacitively loaded strips 
 [Fig.\,\ref{fig:1}(a)]
 at the resonance frequency $f_1$ of the electric dipole resonance and
 in (b) the field-gradient-induced-transparency 
 metamaterial [Fig.\,\ref{fig:1}(b)] at the resonance frequency
 $f_2$ of resonator 2. In the right-hand panels, black (white) 
 corresponds to a large (small) electric field. The outer boxes represent the
 waveguide walls (perfect electric conductors). A section of the metal at
 the center of the I-shaped structure is replaced by a piece of
 dielectric medium 
 in order to tune $f_1$ and $f_2$. The relative
 field strengths in (a) and (b) are not scaled.}
\label{fig:3-2}
\end{center}
\end{figure*}

Figure \ref{fig:3-1} shows the measured transmission spectra 
for the cases in which variable capacitance diodes are connected
(black) and not connected (gray). 
These two cases correspond to the field-gradient-induced-transparency 
metamaterial and the capacitivily loaded strips. 
In the latter case, a simple absorption spectrum is observed, as 
in the case of a Lorentz-type medium. 
In the former case, a transmission window is seen in the
absorption band, as in the case of an EIT-like medium.
Figure \ref{fig:3-2}(a) shows 
the numerically calculated current
and electric field distributions at the resonance
frequency of the capacitively loaded strips shown in Fig.\,\ref{fig:1}(a). 
Figure \ref{fig:3-2}(b) shows the distributions of resonator 2  
in the field-gradient-induced-transparency 
metamaterial shown in Fig.\,\ref{fig:1}(b). 
We used the COMSOL Multiphysics finite element solver. A section of the
metal at the center of the I-shaped structure was replaced with a
piece of dielectric medium so that we can tune the
resonance frequency of resonator 1 to that of resonator 2. 
The current distribution and the electric field distribution enable us
to locate the portions of the structure that act as inductors and 
capacitors. The metal strip in the vertical direction (horizontal
direction) constitutes the inductor for resonator 1 (resonator 2), and
the gap between the neighboring I-shaped structures in the vertical
direction constitutes the capacitor for both resonators. 
It is confirmed that the electric dipole resonance arises at the
resonance frequency in the capacitively loaded strips and that 
the trapped-mode resonance arises 
at the resonance frequency of resonator 2 in the
field-gradient-induced-transparency metamaterial. 
Note that no significant current flows in the vertical
direction in the case of Fig.\,\ref{fig:3-2}(b). 

\begin{figure*}[tb]
\begin{center}
\includegraphics[width=17cm,clip]{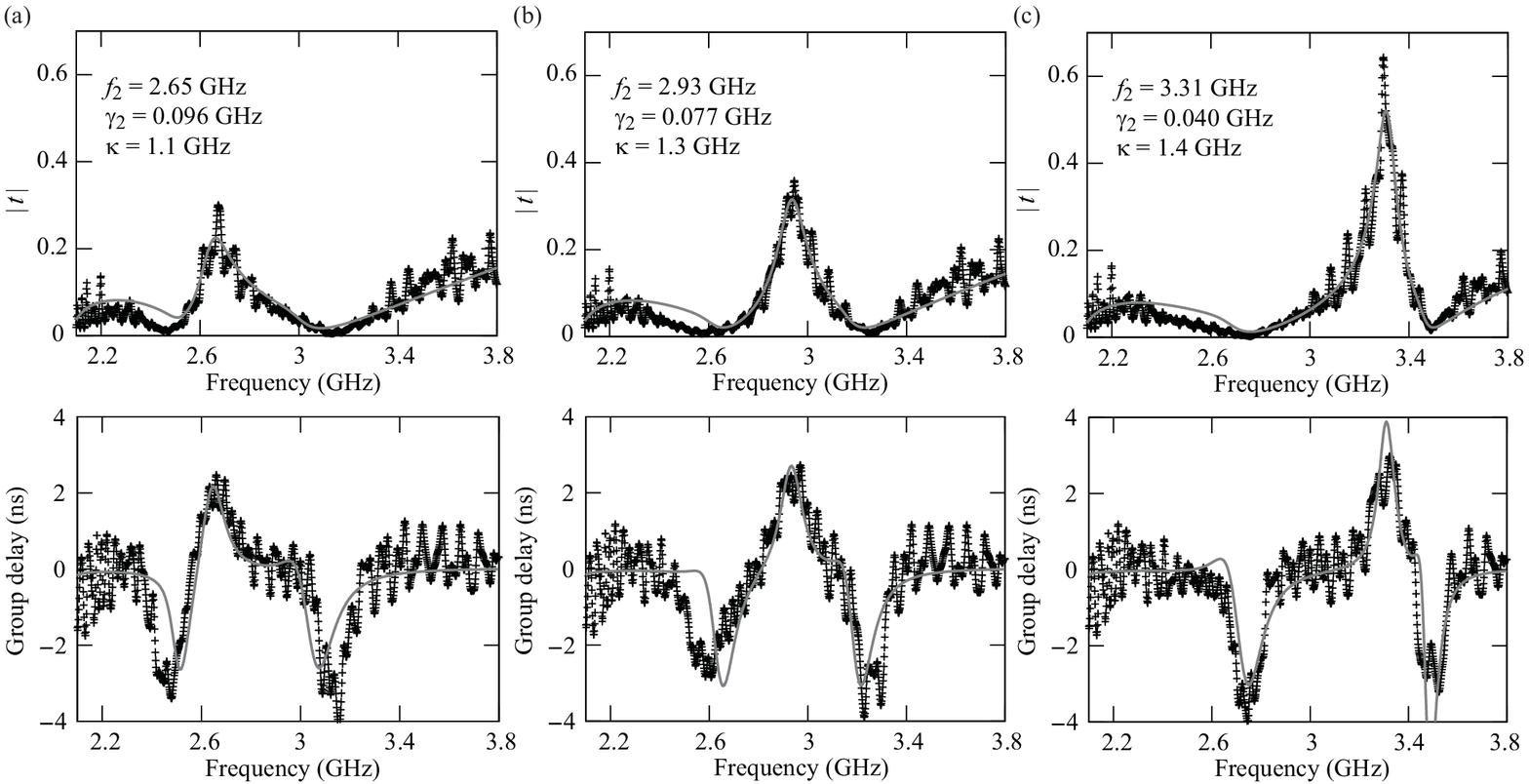}
\caption{Transmissivity (upper row) and group 
delay (lower row) of the fabricated metamaterial as a function of frequency. 
Black crosses and gray solid curves represent the measured data
 and the fitting curves, respectively. The reverse bias voltage on the
 variable capacitance diodes are (a) $0\,\U{V}$, (b) $10.5\,\U{V}$, and 
(c) $30\,\U{V}$.}
\label{fig:4}
\end{center}
\end{figure*}

Figure \ref{fig:4} shows the measured transmissivity and the group delay 
of our metamaterial as a function of frequency 
for three different reverse bias voltages on the variable capacitance diodes. 
The center frequency of the transparency window (low group-velocity band) 
is increased by increasing the reverse bias voltage, i.e., by decreasing 
the capacitance of the diodes. 

Next, we present a quantitative analysis of our metamaterial. 
The charges $q_1$ and $q_2$ in resonators 1 and 2, respectively, are
assumed to satisfy the following equations: 
\begin{align}
&\frac{\dd^2 q_1}{\dd t^2} + (2\pi \gamma_1 ) \frac{\dd q_1}{\dd t} + (2\pi
 f_1)^2 q_1 - (2\pi \kappa)^2 q_2 \nonumber \\
&\hspace{4cm} =
\alpha ^{\prime} \tilde{E} \ee^{-\ii 2 \pi f
 t} + \mathrm{c.c.} ,\\
&\frac{\dd^2 q_2}{\dd t^2} + (2\pi \gamma_2 ) \frac{\dd q_2}{\dd t} + (2\pi
 f_2)^2 q_2 - (2\pi \kappa)^2 q_1 
=
0 ,
\end{align}
where $f$ is the frequency of the incident field, $f_1$ ($f_2$) is the resonance 
frequency of resonator 1 (2), 
$\gamma _1$ ($\gamma _2$) is the loss in resonator 1
(2), $\kappa$ is the coupling factor between resonators 1 and
2, $\tilde{E}$ is the complex amplitude of the incident electric field, 
$\alpha ^{\prime}$ is a constant, and c.c.\ stands for the complex
conjugate of the preceding term.\cite{alzar02}
The electric susceptibility of the metamaterial is given as 
\begin{equation}
\chi \sub{e} (f) = \frac{- \alpha (f^2 +
 \ii \gamma _2 f - f_2 ^2)}{ (f^2 + \ii \gamma _1 f - f_1 ^2 ) (f^2 + \ii
 \gamma _2 f - f_2 ^2 ) - \kappa ^4} \left( \propto
 \frac{\tilde{q_1}}{\tilde{E}} \right) , \label{eq:sus} 
\end{equation}
where $q_1 = \tilde{q_1} \ee ^{-\ii 2\pi ft} + \mathrm{c.c.}$ and
$\alpha$ is a constant.  
Assuming that $f_1
\approx f_2$ and $\gamma_1 \gg \gamma_2$, we find that 
the transmittance at $f_2$, which is the transparent frequency, 
decreases with decreasing $\kappa$, and 
the bandwidth of the transparency window can be approximated 
as $\kappa^2 / f_2$. 
The group velocity at $f_2$ is found to 
decrease with decreasing $\kappa$
until $\kappa^2 / f_2$ reaches approximately $\sqrt{\gamma_1
\gamma_2}$.\cite{zhang_prl08} Therefore,  
we can consider the minimum bandwidth of the transparency window to be
$\sqrt{\gamma_1 \gamma_2}$. 
The transmissivity $t$ of the metamaterial is written as follows:\cite{smith02}
\begin{equation}
t = \frac{4Z_1 Z_2}{[ (Z_2 + Z_1)^2 \ee^{-\ii k_2 d} - (Z_2 - Z_1)^2
 \ee^{\ii k_2 d} ] \ee^{\ii k_1 d}} ,
\end{equation}
where 
$k_1 =(2\pi f /c_0 ) \sqrt{1 - [c_0 / (2wf)]^2}$, 
$k_2 =(2\pi f /c_0 ) \sqrt{1+\chi\sub{e} - [c_0 / (2wf)]^2}$, 
$Z_1 = 2 \pi f \mu_0 / k_1$, 
$Z_2 = 2 \pi f \mu_0 / k_2$,\cite{collin90} 
$w$ is the waveguide width, $\mu_0$ is the permeability of the vacuum,  
and $d$ is the thickness of the metamaterial. 
We adopt $d=1.7\,\U{mm}$ from the physical thickness of the metamaterial. 
It is difficult to define the thickness of a single-layer metamaterial. 
However, in the present case, the change in $d$ has little effect on the following fitting
results. 
We compare $t$ with the measured transmission spectra. 

The fitted curves of $|t|$, and the fitting parameters are shown 
in the upper row of Fig.\,\ref{fig:4}. 
Here, we adopt the values of $f_1 = 2.93\,\U{GHz}$ and $\gamma _1 =
0.20\,\U{GHz}$ derived from the measured transmissivity and group delay 
(not shown) of our metamaterial without the diodes and 
assume that $f_2$, $\gamma _2$, $\kappa$, and $\alpha$ are fitting parameters. 
Although the measured and fitted transmissivities are slightly different
in the lower-frequency region, they agree rather well with each other. 
The group delay $\{ \dd [\arg{(t)}] / \dd f \} / 2\pi$ 
calculated using the obtained fitting parameters is shown in the lower row of 
Fig.\,\ref{fig:4}. 
The measured group delay is in good agreement with the calculated value. 

Based on the above results, the condition whereby $\kappa^2 / f_2 > \gamma_1 >
\gamma_2$ is satisfied. This implies that low-$Q$ resonator 1 is 
strongly coupled to high-$Q$ resonator 2. 
In addition, $f_2 / \gamma _2$ is found to increase 
with the reverse bias voltage. 
This is because the resistance of the variable capacitance diode, which
determines the loss of resonator 2, decreases with 
increasing reverse bias voltage. 
The value $f_2 / \kappa$ is independent of the reverse bias voltage, 
i.e., the coupling strength between resonators 1 and 2 does 
not depend on the resonance frequency of resonator 2. 
A possible reason for the slight difference between the measured and fitted 
values is that we used one layer of the metamaterial in the experiment 
whereas in the theoretical model, we assumed a homogeneous medium. 
The oblique propagation in the waveguide\cite{collin90} 
may also have to be considered. 
 
In order to confirm the $Q$ values of the two types of resonators, 
we performed 
COMSOL simulations and found that $Q_1 = f_1 / \gamma_1 = 9.1$ and $Q_2
= f_2 /
\gamma_2 =71$. The dielectric loss tangent of the substrate was
assumed to be 0.02. The $Q$ values calculated from the fitting
result for the reverse bias voltage of $30\,\U{V}$ were $Q_1
= 15$ and $Q_2 = 83$. The experimental values are slightly larger than
the simulated values for both types of resonators. 
This might be because an actual value of the dielectric
loss tangent of the substrate 
was smaller than 0.02. The ratio $Q_2 / Q_1$ 
for the experiment might be smaller than that for the 
simulation because of the resistance of the diodes.

\section{Simulation of dynamic control of coupling strength}

We performed numerical simulations using COMSOL Multiphysics to 
confirm that the coupling strength between resonators 1 and 2 depends on the field gradient in the
$y$ direction. We calculated the transmissivity of the metamaterial placed in a 
rectangular waveguide. We changed the width of the waveguide in order to control the field
gradient in the $y$ direction of the incident electromagnetic wave and calculated the
transmissivity for several values of the waveguide width. 

\begin{figure}[tb]
\begin{center}
\includegraphics[width=7cm,clip]{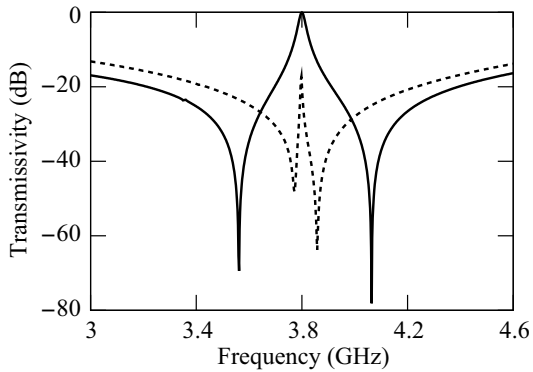}
\caption{Dependence of the transmission spectrum on the waveguide
 width or the field gradient in the $y$ direction of the incident
 electromagnetic wave. The waveguide width for the dashed
curve is 2.2 times larger than that for the solid curve.}
\label{fig:5}
\end{center}
\end{figure}

Figure \ref{fig:5} shows the transmission spectra of the metamaterial shown in
Fig.\,\ref{fig:1}(b) for the TE$_{10}$ mode. 
The geometric parameters of the metamaterial were
$l\sub{a} = 25.1\,\U{mm}$, $l\sub{b} = 18.0\,\U{mm}$, $l\sub{c} =
2.4\,\U{mm}$, $w\sub{a} = 4.0\,\U{mm}$, $w\sub{b} = 4.0\,\U{mm}$, 
and $w\sub{c} = 0.5\,\U{mm}$, and the thickness of the metal was
$0.5\,\U{mm}$. The metal pattern was on a dielectric substrate of 
$0.5\,\U{mm}$ in thickness having a relative permittivity of 4.5. 
In order to tune the resonance frequency of resonator 1 to
that of resonator 2, we replaced part of the metal at the center of the
I-shaped structure 
with a piece of dielectric medium, the length and relative permittivity
of which were $1.8\,\U{mm}$ and
24.5, respectively (see Fig.\,\ref{fig:3-2}). 
The dimensions of the cross section of the waveguide
in the $x$ and $y$ directions
when the solid curve (dashed curve) was obtained were $34.0 \,\U{mm}
$ and $70.0\,\U{mm}$ ($34.0 \,\U{mm}$ and $153.7\,\U{mm}$), respectively. 
We confirmed that the transparency window narrows as the waveguide width
increases. 
In other words, the smaller the field gradient in the $y$ direction of
the incident electromagnetic wave,  
the weaker the coupling strength between resonators 1 and 2.

\section{Summary and discussion}

We investigated a dynamically controllable 
metamaterial that behaves as an EIT-like medium. 
The unit structure of the 
metamaterial is composed of a low-$Q$ resonator (resonator
1) and a high-$Q$ resonator (resonator 2). 
When a plane electromagnetic wave is incident on the metamaterial, 
only resonator 1 can be excited. 
Resonator 1 couples to resonator 2 
when the incident electromagnetic wave has a field-gradient in the
transverse direction, 
such as a Gaussian beam. 
The system is similar to the classical analog of EIT, 
and the metamaterial behaves as an EIT-like medium. 
We described a method by which to control the electromagnetic response
of the metamaterial. 
We experimentally demonstrated dynamic control of the transparent
frequency, i.e., the resonance frequency of resonator 2 
of the metamaterial, using variable capacitance diodes. 
The simulations confirmed that the frequency width of the transparency
window, or 
the coupling strength between resonators 1 and 2, 
can be controlled by the field gradient in the transverse direction of
the incident electromagnetic wave. 

In our experiment, the ratios of the wavelength to the length of the unit 
structure in the $x$ and $y$ directions are 2.3 and 6.5, respectively. 
These ratios are larger than 2, and thus the diffraction can safely be
neglected. 
For normal incidence, in particular, the diffraction does not arise when these
ratios are larger than 1. 
This implies that effective-medium parameters of the metamaterial 
can be defined. That is, 
the metamaterial can be regarded as a homogeneous medium. 
For a larger ratio in the $x$ direction, 
we could reduce the length of the unit structure
in the $x$ direction while preserving the resonance frequencies of
resonators 1 and 2 by reducing $l_1$ and $w_1$.

Although we performed our experiment in the microwave region, 
it is not difficult to conduct similar experiments in higher-frequency
regions, such as the terahertz and optical regions. 
Since the structure of our metamaterial is planar, 
we can fabricate the metamaterial in the optical region by standard 
photolithography techniques. 
We measured the transmission spectrum of the fabricated metamaterial 
in the waveguide, where the beamwidth of the electromagnetic wave is
similar to the wavelength. 
In the optical region, it is possible to obtain a focus spot size 
close to the diffraction limit, and 
hence, it is possible to control the coupling
strength between resonators 1 and 2 in free space. 
We controlled the resonance frequency of resonator 2 by
using variable capacitance diodes. 
The resonance frequency can be controlled by 
using the photoexcitation of carriers in a semiconductor in the 
terahertz region~\cite{chen08} and the phase transitions of 
liquid crystal~\cite{xiao09} and vanadium oxide~\cite{dicken09} 
in the optical region. 

Field-gradient-induced-transparency metamaterials have numerous applications. 
The difference between the capacitively loaded strips 
shown in Fig.\,\ref{fig:1}(a) 
and our metamaterial shown in Fig.\,\ref{fig:1}(b)
is whether or not the I-shaped metal structures are connected to
neighboring structures. 
Therefore, we can dynamically 
switch from a Lorentz-type medium to an EIT-like medium and 
vice versa by controlling the resistance of the connection using 
the photoexcitation of carriers~\cite{padilla06} or Schottky
junction.\cite{chen06} 
This indicates that a dynamic field-gradient-induced-transparency metamaterial could be used as an
amplitude modulator. 
Since the transparency window is narrow and the transparent frequency 
depends on the inductance and capacitance that form resonator 2, 
our metamaterial can be used for
high-sensitivity sensing. 
The metamaterial can also be applied to transverse-mode filters and 
measurements of beamwidth because the transmission spectrum depends on
the field gradient in the transverse direction of the incident electromagnetic wave. 

Field-gradient-induced-transparency bears a resemblance to nuclear
quadrupole resonance. In nuclear quadrupole resonance, 
the energy level of nuclear spin splits 
due to the interaction between the electric quadrupole moment of 
atomic nuclei having a nuclear spin greater than $1/2$ and the 
electric field gradient at the atomic nuclei that is caused 
by electrons. The splitting width of the energy level increases with 
the electric field gradient. In field-gradient-induced transparency, 
the frequency separation between the two dips in the
transmission spectrum increases with the field gradient 
in the transverse direction of the incident electromagnetic waves. 

\begin{acknowledgments}

This research was supported by a Grant-in-Aid for Scientific Research on
Innovative Areas under Grant No.\,22109004 and 
the Global COE program 
\textquotedblleft Photonics and
Electronics 
Science and Engineering\textquotedblright \,at Kyoto University. 
One of the authors (Y.T.) would like to acknowledge the support 
from the Japan Society for the Promotion of
Science.

\end{acknowledgments}

\end{document}